\documentstyle[galley, rotate]{mn}
\onecolumn
\input{epsf}
\def\n{{\noindent}}

\def\inc{{\int_0^{\chi_s}}}
\def\ex{{\int {d^2 {\bf l} \over (2
\pi)^2}~ {\rm P} { \big ( {l\over r(\chi)} \big )} W_2^2(l\theta_0)}}
\def\av{\langle \kappa^2(\theta_0) \rangle}
\def\kmin{{\int_0^{\chi_s} \omega(\chi) d \chi }}
\def\corr{\Big [ \int  {d^2 {\bf l} \over (2
\pi)^2 )} {\rm  P}( { \bf l \over r(\chi) })  W^2(l\theta_0) \exp [ i l
\theta_{12}] \Big ]}
\def\var{ \Big [  \int  {d^2 {\bf l} \over (2
\pi)^2 )} {\rm P} ( { \bf l \over r(\chi) })  W^2(l\theta_0) \Big ] } 
\def\one{\langle \kappa_s^2 \rangle}
\def\two{\langle \kappa_s(\gamma_1) \kappa_s(\gamma_2) \rangle_c}
\def\bi#1{\hbox{\boldmath{$#1$}}}

\title[Weak Lensing in Quasi-linear Regime]
{Probing the Dark Side of Gravitational Clustering:\\
Weak Lensing Statistics at Large Smoothing Angle }

\author[D.Munshi]{Dipak Munshi$^1$\thanks{Present Address:
Institute of Astronomy, Madingley Road, Cambridge CB3 OHA, United Kingdom}\\
Max-Planck-Institut fur Astrophysik,
Karl-Schwarzschild-Str.1, D-85740, Garching, Germany.\\
}
\pagerange{000,000}

\begin{document}

\maketitle

\begin{abstract}
The weak lensing surveys have the potential to probe directly the 
clustering statistics of dark matter in the universe. 
Recent studies have shown that it is
possible to predict analytically the whole probability distribution
function (pdf) and the bias associated with the collapsed objects in the highly
non-linear regime using the hierarchical ansatz. We extend such 
studies to the quasi-linear regime where the hierarchical
ansatz is replaced by the tree-level perturbative calculations to an arbitrary 
order. It is shown how the generating function techniques can be coupled
with the perturbative calculations to compute the complete pdf and the bias 
in the quasi-linear regime for the weak-lensing convergence field.
We study how these quantities depend on the smoothing angle and the source 
red-shift in different realistic cosmological scenarios. We show that it is possible 
to define a reduced convergence whose statistics is similar to underlying 3D 
mass distribution for small smoothing angle but it resembles projected
mass distribution for large smoothing angles. We have also compared our
peturbative results with log-normal model for pdf and bias and found a
good agreement between the two analytical results.
\end{abstract}


\section{Introduction}
The theoretical study of the weak gravitational lensing to probe
 the statistics
of large scale structures in the universe has received much attention
(Mellier 1999, Bartelmann \& Schneider 1999) in recent past.

Based on the earlier works done by several authors (Gunn (1967),
Blandford et al. (1991), Miralda-Escude (1991) and Kaiser (1992)),
 current development 
has focussed mainly on two different directions. Much progress have been 
achieved in the field of numerical simulations of the weak lensing, 
in particular
the construction of shear maps which provide invaluable tools in 
testing the analytical results. The numerical simulations typically employ 
N-body simulations, through which ray tracing experiments are conducted
(Schneider \& Weiss 1988; Jarosszn'ski et al. 1990; Lee \& Paczyn'ski
1990; Jarosszn'ski 1991; Babul \& Lee 1991;  Bartelmann \& Schneider
1991, Blandford et al. 1991). Building on the earlier works of Wambsganns
et al. (1995, 1997, 1998) the most detailed numerical study of lensing 
was done by  Wambsganns, Cen \& Ostriker (1998). Other recent studies
using ray tracing experiments have been conducted by Premadi, Martel
\& Matzner (1998), van Waerbeke, Bernardeau \& Mellier (1998),
Bartelmann et al (1998), Couchman, Barber \& Thomas (1998) and Jain,
Seljak \& White 2000. 

On the analytical front,   Villumsen
(1996), Stebbins (1996), Bernardeau et al. (1997) and Kaiser (1992)
have focussed on the use of perturbative techniques to study weak lensing.
More recently it has been shown that the hierarchical ansatz for the 
higher order correlation functions can be used to describe the 
statistical properties of the weak lensing convergence fields for the 
case of small smoothing angles (Hui 1999, Munshi \& Coles 1999, 
Munshi \& Jain 1999, Valageas 1999a,b, Munshi 2000, Munshi \& Coles 2000a,b).

In this paper we will show that while perturbative calculations have already
been used to compute the lower order moments of convergence field,
(see e.g.Bernardeau, van Waerbeke \& Mellier 1997, van Waerbeke, Bernardeau \& Mellier 1998, for detailed review on perturbative calculations in the context 
of weak lensing surveys see Bernardeau 1999) ; when 
coupled with generating function techniques it can also reproduce the complete
pdf and bias functions associated with the smoothed projected density field.
We also show that these results can simply be obtained by replacing 
the generating function of the tree hierarchy in non-linear regime with its
quasi-linear counterpart. While the non-linear generating function corresponds 
to the three dimensional generating function for the matter correlation 
hierarchy, in the quasi-linear regime it has to be replaced by a generating 
function which represent the smoothed two-dimensional projected
 density field. These generating functions have already been studied in
great detail by Bernardeau (1992, 1994, 1995, 1996) for the case of top-hat
smoothing function. Using the recent analytical results from the study of
statistics of the convergence field for small smoothing angles, we will 
introduce a reduced convergence field $\eta$ and study how its statistical 
properties (including the lower order moments) are related with its 
counterpart used to describe the projected density field.

\section{The Statistics of the Convergence field for Large Smoothing Angles}

The weak lensing convergence $\kappa$ is simply the projected mass distribution and its statistics carries 
 valuable information about the geometry and dynamics of the 
background universe as well as the physics of collision-less clustering.

\begin{equation}
\kappa({\bi \gamma}) = \inc {d\chi}
\omega(\chi)\delta(r(\chi){\bi \gamma})
\end{equation}

\n
Throughout our discussion  we will be placing the sources at a fixed
red-shift (an approximation not too difficult to modify for more
realistic description). The weight function can be expressed as $\omega(\chi)
= 3/2a c^{-2}H_0^2 \Omega_m r(\chi) r(\chi_s - \chi)/ r(
\chi_s)$. Where $\chi_s$ is the comoving radial distance to the source placed
at a red-shift $z_s$. We will be focusing on the quasi-linear regime 
where it is possible to expand the density contrast using a perturbative
series.

\begin{equation}
\kappa^{(1)}({\bi \gamma})+\kappa^{(2)}({\bi \gamma})+\dots = \inc {d\chi}
\omega(\chi)( \delta^{(1)}(r(\chi){\bi \gamma})+\delta^{(2)}(r(\chi){\bi \gamma}) + \dots) 
\end{equation}

\n
It will be useful to work in Fourier domain and the 
 Fourier decomposition of $\delta$ can be written as:

\begin{equation}
\kappa({\bi \gamma}) = \inc {d\chi} \omega(\chi) \int {d^3{\bf k} \over {(2
\pi)}^3} \exp ( i \chi k_{\parallel} + i r \theta k_{\perp} ) \delta_k
\end{equation}

\n
Where we have used $k_{\parallel}$ and $k_{\perp}$ to denote the components
of wave vector ${\bf k}$, parallel and perpendicular to the line of sight
direction ${\bf \gamma}$. In the small angle approximation however, one assumes
that $k_{\perp}$ is much larger compared to $k_{\parallel}$.
 We will denote the 
angle between the line of sight direction ${\bi \gamma}$ and the wave
vector ${\bf k}$, by $\theta$. The smoothing angle will be denoted 
by $\theta_0$. Using the definitions that we have introduced
above we can now express the smoothed
projected two-point correlation function (Limber 1954, Peebles 1980, Kaiser 1992, Kaiser 1998):

\begin{equation}
\langle \kappa({\bi \gamma_1}) \kappa({\bi \gamma_2}) \rangle_c = \inc d {\chi}
{\omega^2(\chi_1) \over r^2(\chi_1)} \int {d^2 {\bf l} \over (2
\pi)^2}~\exp ( \theta l )~ {\rm P} { \big ( {l\over r(\chi)} \big )}
 W_2^2(l\theta_0).
\end{equation}

\n
Where we have introduced a new notation ${\bf l} = r(\chi){\bf
k}_{\perp}$ which denotes the scaled wave vector projected on the
surface of the sky. The average of the two-point correlation function
$\langle \kappa_s^2 \rangle$ 
smoothed over an angle $\theta_0$ with a top-hat smoothing window
$W_2(l\theta_0)$ is useful to quantify the fluctuations in
$\kappa_s$ which is often used to reconstruct the matter power
spectrum $P({\bf k})$ (Jain \& Selzak 1997).

\begin{equation}
\langle \kappa_s^2 \rangle_c = \inc d {\chi  }
{\omega^2(\chi  ) \over r^2(\chi  )} \int {d^2 {\bf l} \over (2
\pi)^2}~ {\rm P} { \big ( {l\over r(\chi)}, \chi \big )} W_2^2(l\theta_0)
\end{equation}

\n
Using the standard perturbative techniques, it is possible to compute
the normalized cumulants or $s_N$ parameters, for the convergence field 
in the large smoothing angle regime.

\subsection{The Probability Distribution Function of $\kappa(\theta_0)$}

To construct the probability distribution function (pdf) it is necessary
 to have
the $s_N$ parameters for the convergence maps computed to an arbitrary order.
These computations have already been done by Bernardeau et al. (1996) and 
Waerbeke et al. (1998).

\begin{eqnarray}
&&\langle \kappa^2(\gamma) \rangle =  C_2[I_{\kappa_{\theta_0}}] \\
&&\langle \kappa^3(\gamma) \rangle = s_3^{\rm PT} C_3[I_{\kappa_{\theta_0}}^2] \\
&&\langle \kappa^N(\gamma) \rangle = s_N^{\rm PT} C_N[I_{\kappa_{\theta_0}}^{N-1}] 
\end{eqnarray}

\n
Where the $s_N$ parameters \footnote{Throughout this paper we will use 
lower case letters for quantities related to  both the convergence $\kappa$ and its
reduced counterpart $\eta$. We will show that the statistics associated 
with the reduced convergence map $\eta({\theta_0}) = {{ \kappa({\theta_0}) - \kappa_{min} } \over
-\kappa_{min}} = 1 + {\kappa({\theta_0}) \over |\kappa_{min}| }$  is very similar to the projected two 
dimensional density fields. The Moments of these fields will be represented by a super-script
${\rm PT}$ to denote their perturbative origin, where as moments associated 
with the convergence map $\kappa$ will be denoted by the superscript $\kappa$. }  correspond to the case of  3D gravitational
 dynamics with 2D perturbations (Bernardeau 1995),

\begin{eqnarray}
&&s_3^{\rm PT} = {36 \over 7} - {3 \over 2} (n+2) \\
&&s_4^{\rm PT} = {2540 \over 49} - 33(n+2) + {21 \over 4} (n+2)^2
\end{eqnarray}

\n
and $C_t$ denotes the line of sight integration.

\begin{equation}
{ C}_t[I^m_{\kappa_{\theta_0}} ] =
\int_0^{\chi_s} { \omega^t(\chi)\over
r^{2(t-1)}(\chi)}I^m_{\kappa_{\theta_0}} d\chi.
\end{equation}

\n
The spectral index $n$ is however same as that of the three dimensional (local)
power spectral index. These results have already been tested using numerical
simulation by Munshi et al. (1999). 
Typically for a degree scale smoothing, we will
use $n=-1.3$. The projection effects are incorporated in the line
 of sight integration. Such a separation of dynamical part and the geometrical
part occurs only when we replace the actual initial power spectra with 
the local power law spectra (Bernardeau, 1995).

\begin{equation}
I_{\kappa_{\theta_0}} = \int { d^2 {\bf l} \over (2 \pi)^2} P\left( {\bf l \over r(\chi)}, \chi \right) W_2(l \theta_0)
\end{equation}

\n
Finally we can write down the normalized cumulants for the convergence field as:

\begin{equation}
s^{\kappa}_N = s_n^{\rm PT} {{C_N[I\kappa_{\theta_0}^{(N-1)}}] \over ~~~~~(C_2[I_{\kappa_{\theta_0}}])^{N-1} }
\end{equation} 

\n 
The generating function approach developed by Balian \& Schaeffer (1989) and
later extended by Bernardeau \& Schaeffer (1992) was used to compute the 
whole hierarchy of $s_N$ parameters for the three dimensional matter distribution
and the projected matter distribution (Bernardeau 1995). 
The generating function is related 
to the vertices which appear in the tree representation of the correlation
hierarchy which appear in the quasi-linear evolution of gravitational clustering.

\begin{equation}
{\cal G}^{\rm PT}(\tau) = 1 - \tau + { \nu_2 \over 2 ! } \tau^2 - { \nu_3 \over
3! } \tau^3 + \dots
\end{equation}

\n
Such a hierarchy also appears in the highly non-linear regime but with 
different tree amplitudes. The generating function for the $s_N$ parameters
can similarly be represented as (Balian \& Schaeffer 1989):

\begin{equation}
\phi^{\rm PT}(y) = \sum_{N=1}^{\infty}(-1)^{N-1} { s^{\rm PT}_N \over N! } y^N
\end{equation}

\n
The function $\phi^{\rm PT}(y)$ satisfies the constraint $s_1 = s_2 = 1$ necessary for
the normalization of the PDF. These two generating function are known to be
related to each other by the following set of equations (Balian \& Schaeffer 1989
, Bernardeau \& Schaeffer 1992):

\begin{eqnarray}
&&\phi^{\rm PT}(y) = y {\cal G}^{\rm PT}(\tau) - { 1 \over 2} y {\tau} { d \over d
\tau}{\cal G}^{\rm PT}(\tau) \\
&&\tau = -y { d \over d\tau} {\cal G}^{\rm PT}(\tau)
\end{eqnarray}

\n
Hence once we know the generating function ${\cal G}(\tau)$ it is possible to
compute the whole hierarchy of the projected $s_N$ parameters. It was shown
that the generating function takes a particularly simple form (Bernardeau 1995):

\begin{equation} 
{\cal G}^{\rm PT}(\tau) = \Big ( 1 - {\tau \over \kappa_a} \Big )^{-\kappa_a},
\end{equation}

\n
with 
$\kappa_a = {(\sqrt{13} -1)/ 2}$
for 2D perturbative dynamics of gravitational clustering in a 3D universe.
This particular result hold for unsmoothed density field but it is 
possible to incorporate the effect of smoothing into account if we 
assume that the smoothing window is a top-hat function (Bernardeau 1994, 
Bernardeau 1995).

\begin{equation}
{\cal G}_s^{\rm PT}(\tau)={\cal G}^{\rm PT}\left[\tau\,
{\sigma\left(R_0[1+{\cal G}^{\rm PT}(\tau)]^{1/2}\right)\over
\sigma(R_0)}\right].
\end{equation}

\n
For the variance $\sigma$ at length scale $R_0$ will use the local power spectral
index $n$ so that  $\sigma(R_0) \propto R_0^{-(n+3)/2}$.

\n
The VPF and the PDF can be related to each other by the following equation
(Balian \& Schaeffer 1989):
\begin{equation}
P^{\rm PT}(\delta) = \int_{-i\infty}^{i\infty} { dy \over 2 \pi i} \exp \Big [ {(
1 + \delta )y - \phi^{\rm PT}(y)  \over \bar \xi_2} \Big ] \label{pdf}.
\end{equation}

\n
The above expressions are clearly suitable for the density field and 
similar calculations can be done for convergence maps and the generating
functions for the cumulants of the convergence maps can then directly be related to
its counterpart for the density field. The details of such calculations 
can be found
in Munshi \& Jain (1999) and Valageas(1999b). Our notations follow
that of Munshi \& Jain (1999) more closely. These studies clarified that
it is useful to define a reduced convergence map $\eta$ smoothed with an
angle $\theta_0$;

\begin{equation}
\eta({\theta_0}) = {{ \kappa({\theta_0}) - \kappa_{min} } \over
-\kappa_{min}} = 1 + {\kappa({\theta_0}) \over |\kappa_{min}| } ,
\end{equation}

\n
where the minimum value of the convergence along any line of sight direction
can be defined by the following equation:
\begin{equation}
\kappa_{min} = - \int_0^{\chi_s} d \chi \omega(\chi) .
\end{equation}

\n
Under certain approximation $\eta$ will have the exactly same one-point
statistics as the density field $1 + \delta$.

\begin{equation}
\Phi_{\eta}(y) = \phi^{PT}(y) \\
\end{equation}

\begin{figure}
\protect\centerline{
 \epsfysize = 3.5truein
 \epsfbox[23 393 338 717]
 {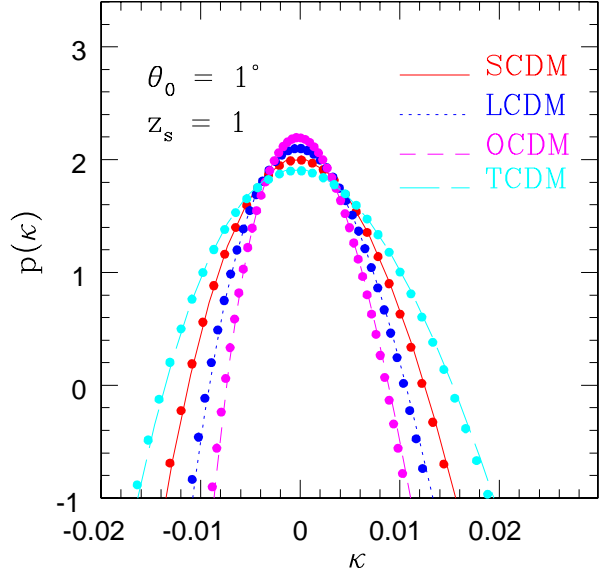} }
 \caption{The Probability distribution function associated with 
different realistic scenarios.
The smoothing angle is 
fixed at $\theta_0= 1^{\circ}$. The source is placed at $z_s =1$ for each of these models.
 The slope of correlation function $\gamma = 1.7$ was assumed. Dots represent results from 
the log-normal fitting form with same variance and $k_{min}$.}
\end{figure}

\n
Notice that the quantity $k_{min}$ does not depend on the smoothing
angle $\theta_0$ and hence will only depend on the source red-shift $z_s$ and 
the definition of reduced convergence remains same both in the quasi-linear
and the highly non-linear regime and is not connected with the convergence 
or divergence of the perturbative series.

In general the expression connecting the cumulant generator for $\eta$ 
,i.e. $\Phi_{\eta}$ can be related to $\phi$ by the following approximation.

\begin{equation}
\Phi_{\eta} (y) = { 1 \over [\kmin]} \int_0^{\chi_s} d \chi 
\Big [{ r^2(\chi) \over \kmin}{ \av \over \ex }\Big ] \phi^{\rm PT} \Big [
\omega(\chi) \kmin { \ex \over r^2(\chi) \av}   \Big ] 
\end{equation}

Clearly this expression does not depend on whether we are in the quasi-linear 
regime or in the highly non-linear regime. Only changes which are to 
be made is that of $\phi$. In quasi-linear regime the quasi-linear phi
can be computed from quasi-linear generating function for tree vertices
defined earlier and in highly non-linear regime it needs to replaced
with its non-linear counterpart. The simplification mentioned above can 
be achieved by replacing the red-shift dependence of different integrands
with their values at a median red-shift. Numerical computations seems to
suggest that such an approximation works remarkably well (Valageas 1999b, 
Munshi \& Jain 1999a). 

It is possible to have different approximate asymptotic expressions 
for different limiting values of $\eta$. However here we have performed 
a direct numerical
integration and results for different cosmologies and various smoothing angles are presented in the 
Figure-1 and Figure-2. The different curves represent various versions of (standard, lambda, omega 
and $\tau$) cold dark matter models. The details of various parameters
charecterising these dark matter models can be found in Munshi \& Jain
(1999a). 

\n
As in the case of quasi-linear regime we now get:

\begin{eqnarray}
&&s_3^{\kappa} =s_3^{\rm PT}/ \kappa_{min}  \\
&&s_4^{\kappa} =s_4^{\rm PT}/ \kappa_{min}^2  \\
&&s_N^{\kappa} = s_N^{\rm PT}/ \kappa_{min}^{N-2}.
\end{eqnarray}

\n
Although these expressions are very similar to 
their small angle counterparts (Munshi \& Jain 1999a,
Valageas 1999), however it is not very difficult to
notice that there is a sharp difference between the two. For larger 
smoothing angle the statistics of reduced convergence map follows 
that of projected density field which in turn is related to the results
associated with the two dimensional
perturbations in a three dimensional universe. However in case of
very small smoothing angles when we are probing the non-linear 
regime, the statistics of the convergence maps is that of full three
 dimensional density
field. The intermediate regime is more difficult to model and it 
interpolates these two regimes in a smooth manner. Numerical investigations
of $s_3^{\kappa}$ was done by Gaztanaga \& Bernardeau (1998) for SCDM model
for various smoothing angle. We find a good agreement between
our results and their numerical measurements for angular scales
as small as $\theta = 15'$. Which means that 
our analytical results for pdf and bias (based on tree-level
calculations), may even be valid for 
much smaller angular scales. A detailed
comparison however will require a large simulation patch and is not available
at present.


\begin{figure}
\protect\centerline{
 \epsfysize = 3.5truein
 \epsfbox[20 414 591 717]
 {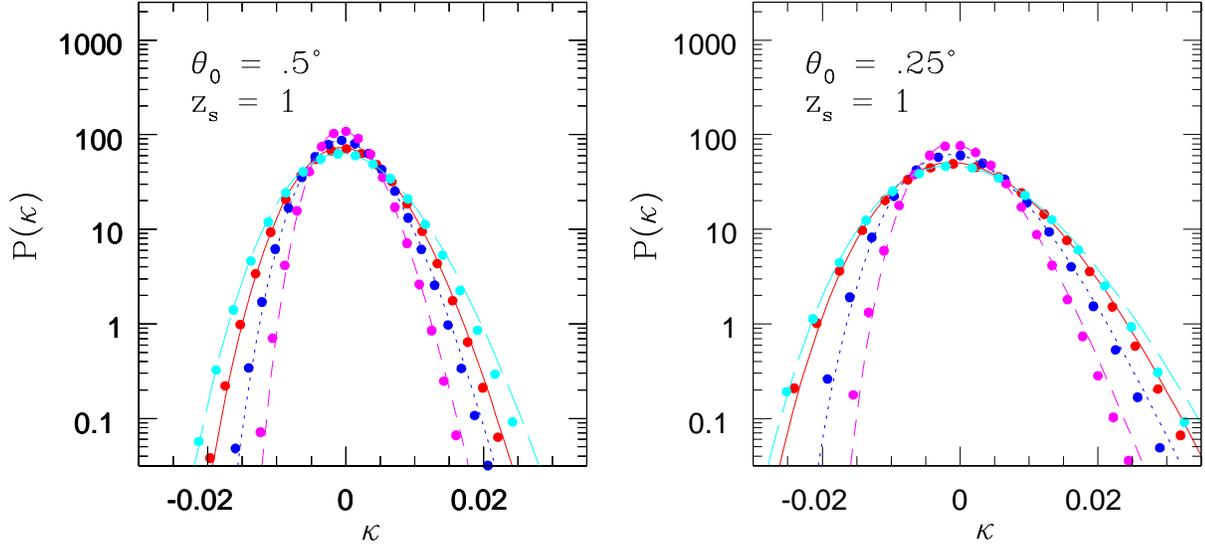} }
 \caption{The Probability distribution function associated with 
different realistic scenarios.
The smoothing angle is 
fixed at $\theta_0= .5^{\circ}$ in the left panel and at $\theta_0= .25^{\circ}$ in the right panel . The source is placed at $z_s =1$ for each of these models. The slope of correlation function $\gamma = 1.7$ was assumed. Dots represent results from the log-normal fitting formula with same variance and $k_{min}$.}
\end{figure}

\subsection{The bias associated with the convergence field $\kappa(\theta_0)$}

While it is clear that one-point studies of convergence field is interesting
it was shown by Munshi (2001) that such results can also be obtained for 
the case of two-point quantities such as bias associated with the 
convergence map $b(\kappa)$ and its moment, the cumulant correlators $c_{pq}$
(see Munshi, Melott \& Coles 1999 for detailed descriptions of generalized
cumulant correlators based on various hierarchical models).

\begin{equation}
p(\kappa_1,\kappa_2)d\kappa_1 d\kappa_2 = p(\kappa_1)p(\kappa_2)(1+b(\kappa_1)
\xi^{\kappa}_{12}b(\kappa_2) ) d\kappa_1 d\kappa_2
\end{equation}

\n
We have used shorthand notations $\kappa_1 = \kappa(\gamma_1)$, $\kappa_2 =
\kappa(\gamma_2)$ and 
$\xi^{\kappa}_{12} = \langle \kappa(\gamma_1) \kappa(\gamma_2)
\rangle$ to simplify the above expression. A similar expression can be 
obtained for the reduced convergence map $\eta$.
 
Our studies in the highly non-linear regime have shown that the hierarchical 
ansatz can be used to understand how the `hot-spots' in convergence 
maps are correlated and how such correlations can be used to study
the similar correlation for the underlying matter distribution and 
hence the bias associated with the collapsed objects. We will show that 
a similar study is also possible in the quasi-linear regime with 
appropriate change in the generating function as was the case for
non-linear regime. 

The normalized cumulant correlators or $c_{pq}^{\rm PT}$ parameters have been
 calculated for dark matter density 
perturbations using direct perturbative calculations for low orders
and for arbitrary orders using the generating function techniques. 
We can define these parameters by the following relation

\begin{equation}
\langle \kappa_s^p({\bi \gamma_1}) \kappa_s^q({\bi \gamma_2})\rangle_c  =
= c_{pq}^{\kappa} \langle
\kappa_s^2 \rangle_c^{(p+q-2)} \langle \kappa_s({\bi \gamma_1}) \kappa_s({\bi \gamma_2}) \rangle.
\end{equation}

\n
These parameters were computed by Bernardeau(1996) for the density field
 and were found to obey a
generic factorization rule $c_{pq} = c_{p1}c_{q1}$ for the case of 3D
perturbations. Similar computations were later generalized for the case
of 2D perturbations. 

\begin{eqnarray}
&&c_{21}^{\rm PT} =  {24 \over 7} - {1 \over 2} (n+2), \\
&&c_{31}^{\rm PT} = {1473 \over 49} -{195 \over 14} (n+2) + {3 \over 2} (n+2)^2.
\end{eqnarray}

\n
For projected fields we will get a prefactor resulting from the
 line of sight integration as was the case
for the one-point cumulants.

\begin{equation}
\langle \kappa_s^p({\bi \gamma_1}) \kappa_s^q({\bi \gamma_2})\rangle_c  =
 c_{pq}^{\rm PT}{\cal C}_{p+q}[I^{(p+q-2)}_{\kappa_{\theta_0}} I_{\kappa_{\theta_{12}}}]=c^{\rm PT}_{p1}c^{\rm PT}_{q1} {\cal C}_{p+q}[I^{(p+q-2)}_{\kappa_{\theta_0}} I_{\kappa_{\theta_{12}}}]
\end{equation}

\n
where we have introduced a new quantity $I_{\kappa_{\theta_{12}}}$ which 
incorporates the line of sight integration effects for the two-point quantities,
such as the cumulant correlators and bias which we are considering here.

\begin{equation}
I_{\kappa_{\theta_{12}}}  \equiv  \int
 \frac{d^2\bf l}{(2\pi)^2} P \left( {l \over r(\chi)} \right)
W_2^2(l \theta_0) \exp ( l \theta_{12}).
\end{equation}.

\n
It can be shown that the role of the generating function for the normalized 
cumulant correlators for the (unsmoothed) density field 
$1+\delta$ $c^{\rm PT}_{p1}$ is played by the quantity $\tau$ itself. 

\begin{equation}
\beta^{\rm PT}(y) = \sum_{p=1}^{\infty} {c^{\rm PT}_{p1} \over p!} y^p
\end{equation}

\n
In a very similar way we can define the generating function $c_{pq}$ for 
the convergence map $\kappa$ by the following relation:

\begin{equation}
\beta_{\kappa}(y_1, y_2) = \sum_{p,q}^{\infty} {c^{\kappa}_{pq} \over p! q!} y_1^p y_2^q = 
\sum_{p}^{\infty} {c^{\kappa}_{p1} \over p!} y_1^p \sum_{q}^{\infty} {c^{\kappa}_{q1}
\over q!} y_2^q  = 
\beta_{\kappa}(y_1) \beta_{\kappa}(y_2) 
\end{equation}

\n
However such a factorization is not possible for the projected density fields, 
without some simplifying assumptions and in general we have to use 
the generating function for $c^{\rm PT}_{pq}$ which we denote as 
$\beta_{\kappa}$. Using relations we have already presented above we
can finally write:

\begin{eqnarray}
\beta_{\kappa}(y_1, y_2) &=& \sum_{p,q}^{\infty} {c_{pq}^{\rm PT} \over p! q! } { 1 \over 
\langle \kappa_s^2 \rangle^{p+q -2}} { 1 \over \langle
\kappa_s(\gamma_1) \kappa_s (\gamma_2) \rangle } \nonumber \\
&& \times  \int_0^{\chi_s} d\chi
{ \omega^{p+q} \over r^{2(p+q -1)} } \Big [ \int  {d^2 {\bf l} \over (2
\pi)^2 )} P( { \bf l \over r(\chi) })  W^2(l\theta_0) \exp [ i l
\theta_{12}] \Big ]
\Big [ \int  {d^2 {\bf l} \over (2
\pi)^2 )}  P( { \bf l \over r(\chi) })  W^2(l\theta_0) \Big ]^{p+q-2} y_1^p y_2^q.
\end{eqnarray}

\n
As was the case with one point cumulants we can use the reduced convergence
$\eta(\theta_0)$ to simplify the analysis because as pointed out earlier
the statistics of reduced convergence map is very similar to that of 
the underlying projected density field.

\begin{eqnarray}
\beta_{\eta}(y_1, y_2) &=& \inc \omega^2(\chi) d\chi  {{\left ( \omega^2(\chi) \over
r^2(\chi) \right )} \corr \over \two } { \one^2 \over
{\left ( \omega(\chi) \over
r^2(\chi) \right )^2} \var^2 \Big [ \kmin \Big ]^2
} \nonumber  \\ &&
\times ~~\beta^{\rm PT} \Big ( { y_1 \over \one} {\omega(\chi) \over r^2
(\chi)} \var \kmin  \Big ) \nonumber \\ &&
\times ~~ \beta^{\rm PT} \Big ( { y_2 \over \one} {\omega(\chi) \over r^2
(\chi)} \var  \kmin \Big ).
\end{eqnarray}

\n
If we use the same techniques which we have used to simplify the integrals 
associated with one the point cumulants (see Munshi \& Jain 1999a, and Munshi 2000)
 where we replaced the red-shift dependence
in integrands with their values at a median red-shift we can write:

\begin{equation}
\beta_{\eta}(y_1, y_2) = \beta_{\eta}(y_1) \beta_{\eta}(y_2) =
\beta^{\rm PT}(y_1) \beta^{\rm PT}(y_2) 
\end{equation}

\n
This is exactly the same result which we have obtained for small smoothing
angle, however results we have derived here are valid for large smoothing
angles and hence the generating function $\beta^{\rm PT}(y)$
 corresponds to the case
of perturbative regime. If we neglect the effect of 
smoothing the generating function $\beta^{\rm PT}(y)$ is simply the 
function $\tau(y)$ (Bernardeau 1996) however for smoothed convergence
maps we can write (Bernardeau 1996):

\begin{equation}
\beta^{\rm PT}(y) = \tau(y) {\sigma(R_0) \over \sigma\left(R_0[1+{\cal G}^{\rm PT}(\tau)]^{1/2}\right)}.
\end{equation}

Finally the bias $b(\eta)$ can be expressed as:

\begin{equation}
b(\eta) = \int_{-i\infty}^{i\infty} { dy \over 2 \pi i} ~\beta(y) \exp \Big [ { \eta y - \phi^{\rm PT}(y)  \over \bar \xi_2} \Big ] /  \int_{-i\infty}^{i\infty} { dy \over 2 \pi i} ~ \exp \Big [ { \eta y - \phi^{\rm PT}(y)  \over \bar \xi_2} \Big ]  \label{bias}.
\end{equation}

\n
The generating function for $\eta$ and projected density field 
are same under certain simplifying approximation, it shows that the
 bias associated with the reduced convergence
is very similar to the statistics of the projected density field.
This is in direct contrast with what was obtained using the nonlinear 
theory for small smoothing angle where the statistics of hot-spots 
in projected map are more closely related to the underlying three dimensional
density field and not to the projected density field. Finally the 
bias associated with the convergence $\kappa$ can be expressed
by the following equation.

\begin{equation}
b_{\kappa}(\kappa) = {b_{\eta}(\eta) \over {k_{min}}}.
\end{equation}

\begin{figure}
\protect\centerline{
 \epsfysize = 3.5truein
 \epsfbox[23 393 338 717]
 {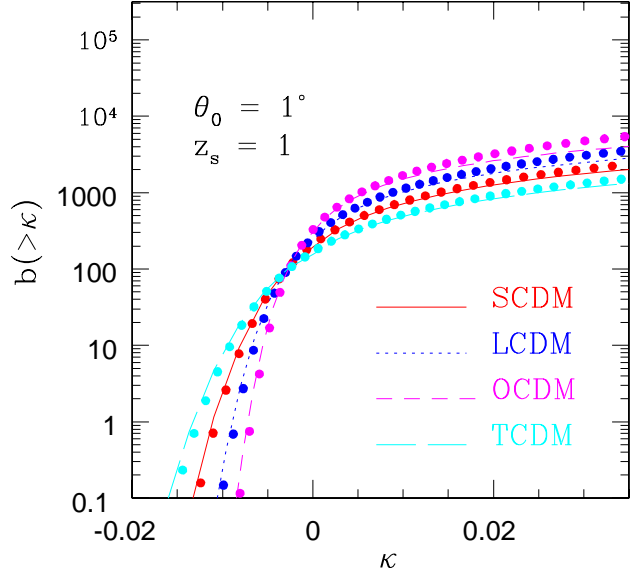} }
 \caption{Bias associated with different realistic scenarios.
The smoothing angle is 
fixed at $\theta_0= 1^{\circ}$. The source plane is placed at $z_s =1$ for each of these models. The slope of the correlation function $\gamma = 1.7$ was assumed. Dots represent predictions from log-normal model.}
\end{figure}

\n
Clearly it is possible to integrate the exact equation which represents
joint moments exactly and hence compute the exact bias function for large
smoothing angle from $\beta(y)$, but it was found by Munshi(2000) that
 simplification of replacing all the integrals by their approximate values,
works remarkably well, when tested against numerical simulations. We will
adopt the same technique for this study. Result of such numerical integration
for different cosmologies are presented in Figure-3 and Figure-4.  Instead of plotting
the differential bias function we have plotted its integral part which 
simply represent the bias associated with spots which cross certain threshold
in reduced convergence. Such a bias is easier to measure from numerical maps.
We find that the bias predictions based on log-normal models also match 
perturbative calculations reasonably well.

While in the non-linear regime it is possible to extend such studies to even 
higher order where one gets the complete picture for the many-point 
statistics of over-dense cells, similar analysis is not possible in 
quasi-linear regime as the tree-vertices $\nu_n$ defined above are 
dependent on shape factors.

Different sources of noise affect the measurement of one-point quantities 
such as pdf and two-point quantities such as bias from observational data or
 numerical simulations.
While for small angular scales the major concern is always the 
noise due to intrinsic ellipticities of galaxies at larger smoothing
angle it is finite volume corrections which starts playing the major role.
It manifests itself mainly on large $\kappa$ tail of the pdf and bias
and depends on the presence or absence of a rare cluster in the catalog.
The large $\kappa$ tail generally shows large fluctuations before showing
an abrupt cutoff for a certain threshold beyond which there are no more
over-dense cells in the catalog. Similar effects are observed in the
determination of bias too.

\begin{figure}
\protect\centerline{
 \epsfysize = 3.5truein
 \epsfbox[20 414 591 717]
 {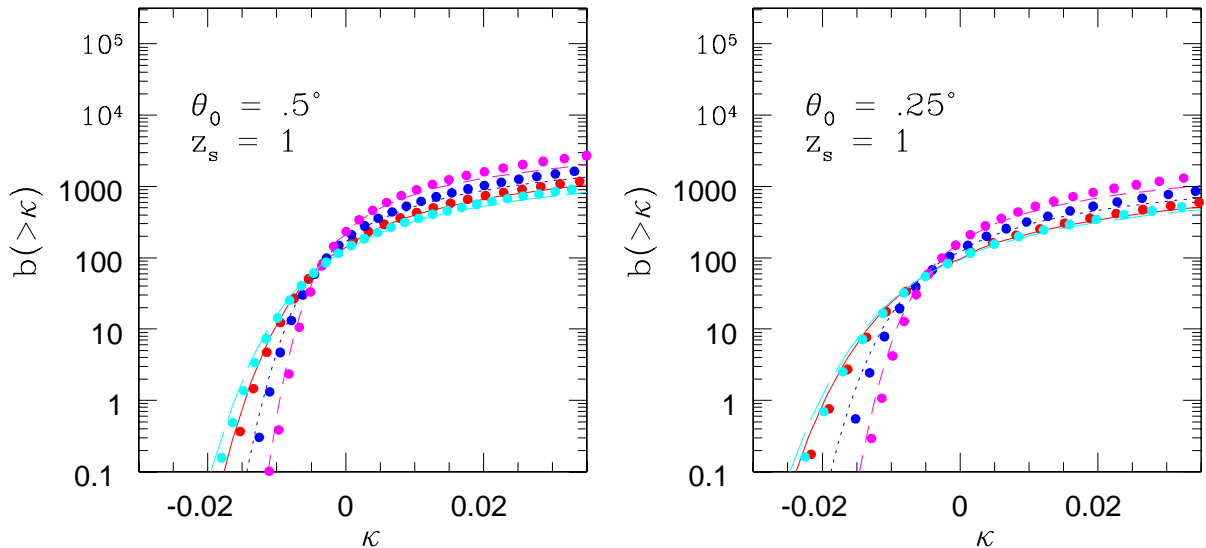} }
 \caption{The bias function associated with 
different realistic scenarios.
The smoothing angle is 
fixed at $\theta_0= .5^{\circ}$ in the left panel and at $\theta_0= .25^{\circ}$ in the right panel . The source plane is placed at $z_s =1$ for each of these models. The slope of correlation function $\gamma = 1.7$ was assumed. Dots represent results from the log-normal fitting formula with the same variance and $k_{min}$.}
\end{figure}

\section{Discussion}

Perturbative calculations have already been used to compute the lower
order hierarchy of $s_N$ parameters associated with the convergence maps
in quasi-linear regime. On the other hand hierarchical ansatz was
shown to be useful in highly non-linear regime. In this particular
paper we have shown how the recent development in highly non-linear regime
can also be used to study the quasi-linear regime by introducing a 
reduced convergence field to construct the full pdf and bias associated 
with convergence maps in the quasi-linear regime. 
We have shown that  a reduced convergence
field in fact is related directly with the two-dimensional projected
density field. It is known from earlier studies that the generating 
function corresponding to the two-dimensional density field is connected with
the two dimensional spherical collapse. Which indicates that such a
generating function now can be used to predict the pdf and bias of
smooth convergence maps in the quasi-linear regime. A quick comparison
with corresponding results in highly non-linear regime shows that while
statistics of projected density for small smoothing is intimately 
related with the full three dimensional density field at larger 
smoothing angle it is related to the projected two dimensional 
density field.

From observational point of view our results will have direct significance.
While most recent observational studies cover rather small patches 
of the sky necessary for quasi-linear approximations to be valid,
 it is hoped that 
larger and larger patches of the sky will be covered in future, 
where one can apply perturbative techniques. At smaller angular 
scales it is well known that noise due to intrinsic ellipticity of 
galaxies will play a major role in determination of the cosmological
parameters, however such noise will be less significant for 
studies at larger angular scales where our results will have direct
relevance. Although signal variance will start declining with 
larger and larger smoothing angle and direct determination of 
weak lensing signal at degree scale might seem difficult at present,
in the intermediate length scale $\theta_0 = 15'$ where the hierarchical
ansatz starts to break down our analysis using perturbative techniques
will provide much needed theoretical insight.

Most of the presently available numerical maps for cosmic shear 
are generated from high resolution numerical simulation and are 
generally correspond to few degree patches in the sky which are 
suitable for testing analytical predictions for smaller smoothing
angles, and recent comparison of various statistics were found to 
be very accurately reproduced by non-linear theory. Such studies
for larger smoothing angles will have to wait till much larger
numerical maps are available. There have been some attempts in this
direction using higher order Lagrangian theories which replaces
the exact gravitational dynamics. Such models although less accurate
can reproduce the basic features of convergence maps in larger smoothing 
angles and are definitely very cost effective.
 For reproducing the statistics of convergence maps made using
higher order Lagrangian perturbation theory we just need to change 
the exact generating function which we have used in this study by 
corresponding smoothed generating function for perturbative
Lagrangian dynamics for the case of projected perturbations and it 
can then be used to compare with numerical studies. It can be shown
that the frozen flow approximation as proposed by Matarrese et al(1992).
will produce a log-normal pdf in the quasi-linear regime (Munshi, Sahni \& Starobinsky 1994), 
hence such an approximation can also be used to generate convergence
maps. 

Results presented here can be generalized very easily for the case 
of projected galaxy survey. A detailed study will be presented elsewhere.

\section*{Acknowledgment}
I was supported by a fellowship from the Humboldt foundation at the MPA
when this work was completed. It is pleasure for me to acknowledge
many helpful discussions with  Francis Bernardeau, Bhubnesh Jain,  
Patrick Valageas and Peter Coles. The complex integration routine I
 have used to generate $p(\eta)$ and
$b(>\eta)$ were made available to me by Francis Bernardeau. I am
grateful to him for his help.

\section*{Appendix: Log-normal Distribution}

Lognormal distribution provides a phenomological description of
one -point probability ditribution function (see Coles \& Jones 1991)
and its generalisation to compute the bias function associated
with overdense region (Taruya et al. 2001). Although inherently local 
in nature we found that it can also be used as a good fitting function
for both one-point probability distribution function 
(see Bernardeau \& Kofman 1995 for similar comparison in 3D for density PDF) and the bias 
function associated with the convergence maps. For the projected 
density contrast $\delta$ (which is related to the
reduced cnvergence field $\eta = 1 + \delta$)  we can write the pdf as:

\begin{equation}
P(\delta)d\delta = {1 \over \sqrt{2 \pi \Sigma^2}} \exp \Big[-{\Lambda^2\over 2 \Sigma} \Big] { 1 \over (1+\delta)}
d\delta.
\end{equation}

\n
Similarly the joint probability distribution function can be written as (Kayo et al.2001):

\begin{equation}
P(\delta_1,\delta_2)d\delta_1d\delta_2 = {1 \over 2 \pi \sqrt{(\Sigma^2 - X^2)}} \exp \Big [ -{\Sigma(\Lambda_1^2+\Lambda_2^2) 
-2X\Lambda_1\Lambda_2 \over 2(\Sigma^2-X^2)}\Big ]  {1 \over (1+\delta_1)}{1 \over ( 1+ \delta_2)}
d\delta_1d\delta_2
\end{equation}

\n
Where we have introduced the following notations:

\begin{equation}
X= \ln(1+\xi_{nl})~, ~~~\Sigma=(1+\sigma_{nl})~, ~~~\Lambda_i = \ln\big [(1+\delta_i) \sqrt{(1+\sigma_{nl})} \big ].
\end{equation}

\n
$\xi_{nl}$ and $\xi_{lin}$ denotes the non-linear and linear correlation functions and can be computed
using suitable modelling of non-linear evolution e.g.  Peacock \& Dodds (1996). To compute the
one-point and two-point pdf's for $\kappa$ from above expressions we have to apply the transformation
$\eta \rightarrow \kappa$ (which involve computation of $\kappa_{min}$) discussed before.
 A detailed comparison of bias from log-normal model and from
hierarchical ansatz for the case of density field will be presented elsewhere.


\begin{thebibliography}{}
\bibitem{} Babul, A., \& Lee, M.H., 1991, MNRAS, 250, 407
\bibitem{BaSa} Balian R., Schaeffer R., 1989, A\& A, 220, 1
\bibitem{} Bartelmann M., Huss H., Colberg J.M., Jenkins A., Pearce F.R., 1998,A\& A, 330, 1 
\bibitem{} Bartelmann, M. \& Schneider, P., 1991, A\& A, 248, 353
\bibitem{B92} Bernardeau F., 1992, ApJ, 392, 1
\bibitem{}  Bernardeau F., 1994, A\& A, 291, 697
\bibitem{}  Bernardeau F., 1995, A\& A, 291, 697
\bibitem{}  Bernardeau F., 1996, A\& A, 312, 11
\bibitem{} Bernardeau F., \& Kofman L., 1995, ApJ, 443, 479
\bibitem{} Bernardeau F., Schaeffer R., 1992, A\& A, 255, 1
\bibitem{} Bernardeau F., Schaeffer R., 1999, A\& A, 349, 697
\bibitem{} Bernardeau F., van Waerbeke L., Mellier Y., 1997, A\& A,
322, 1
\bibitem{} Blandford R.D., Saust A.B., Brainerd T.G., Villumsen
J.V., 1991, MNRAS, 251, 600
\bibitem{} Coles P., \& Jones B., 1991, MNRAS, 248, 1
MNRAS, 308, 180
MNRAS, 307, 537
\bibitem{} Gunn, J.E., 1967, ApJ, 147, 61
ApJ, 374, L1
\bibitem{} Hui, L., 1999, ApJL, 519, 9 
\bibitem{} Jain, B., \& Seljak, U., 1997, ApJ, 484, 560
\bibitem{} Jain, B., Seljak, U., White, S.D.M., 2000, ApJ, 530, 547
\bibitem{} Jaroszyn'ski, M., Park, C., Paczynski, B., \& Gott, J.R.,
1990, ApJ, 365, 22
\bibitem{} Jaroszyn'ski, M. 1991, MNRAS, 249, 430
\bibitem{} Kaiser, N., 1992, ApJ, 388, 272
\bibitem{} Kaiser, N., 1998, ApJ, 498, 26
MNRAS, 303, 188
MNRAS, 307, 529
\bibitem{} Limber D.N., 1954, ApJ, 119, 665
\bibitem{} Lee, M.H., \& Paczyn'ski B., 1990, ApJ, 357, 32
\bibitem{} Matarrese S., Francesco L., Moscardini L., \& Saez D., 1992, MNRAS, 259, 437
\bibitem{} Miralda-Escud\'{e} J., 1991, ApJ, 380, 1
\bibitem{} Munshi D., Sahni V., Starobinsky, A. A., 1994, ApJ, 436, 517
\bibitem{} Munshi D., Bernardeau F., Melott A.L., Schaeffer R.,
1999, MNRAS, 303, 433
\bibitem{} Munshi D., Melott A.L., Coles P., 1999, MNRAS, 311, 149
\bibitem{} Munshi D. \& , Coles P., 2000a, MNRAS, 313, 148
\bibitem{} Munshi D. \& , Coles P., 2000b, MNRAS, submitted
\bibitem{} Munshi D. \& , Jain B.,  2000, MNRAS,318,109
\bibitem{} Munshi D. \& , Jain B.,  2001, MNRAS,322,107
\bibitem{} Munshi D., 2000, MNRAS, 318, 1
\bibitem{} Mellier Y., 1999, ARAA, 37, 127
ApJ, 466, 604
\bibitem{} Peacock J.A., \& Dodds, S.J., 1996, MNRAS, 280, L19
\bibitem{} Premadi P., Martel H., Matzner R., 1998, ApJ, 493, 10
\bibitem{} Peebles P.J.E., 1980, {\em The Large Scale Structure of the
Universe}. Princeton University Press, Princeton
\bibitem{} Schneider P., \& Weiss, A., 1988, ApJ, 330,1
Melott A.L., 1998, ApJ, 496, 586
\bibitem{} Stebbins A., 1996, astro-ph/9609149
\bibitem{} Taruya A., Takada M., Hamana T., Kayo I. \& Futamase T., 2002, astro-ph/0202090
\bibitem{} Valageas P., 1999a, 2000, A\&A, 354, 767
\bibitem{} Valageas P., 1999b, 2000, A\&A, 356, 771
\bibitem{} van Waerbeke, L., Bernardeau, F., Mellier, Y., 1999,
A\&A, 342, 15
\bibitem{} Villumsen J.V., 1996, MNRAS, 281, 369
\bibitem{} Wambsganss J., Cen, R., \& Ostriker, J.P., 1998, ApJ, 494,
298
\bibitem{} Wambsganss J., Cen, R., Xu, G. \& Ostriker, J.P., 1997, ApJ, 494,
29
\bibitem{} Wambsganss J., Cen, R., Ostriker, J.P. \& Turner, E.L.,
1995, Science, 268, 274
\end{thebibliography}
\end{document}